\documentclass[12pt]{article}
\usepackage{psfrag}
\usepackage{graphicx}
\usepackage{graphics}
\usepackage{epsfig}
\newcommand{\As}{A\!\!\!/}
\newcommand{\ks}{k\!\!\!/}
\newcommand{\ps}{p\!\!\!/}

\newcommand{\ssl}{s\!\!\!/}
\date{}
\begin{document}
\baselineskip=18.6pt plus 0.2pt minus 0.1pt \makeatletter


\title{\vspace{-3cm}
\hfill\parbox{4cm}{\normalsize \emph{}}\\
 \vspace{1cm}
{Spin effects in laser-assisted semirelativistic excitation of atomic hydrogen by electronic impact}}
 \vspace{2cm}
\author{S. Taj$^1$, B. Manaut$^{1,2}$\thanks{{\tt b.manaut@usms.ma}}, M. El Idrissi$^1$, Y. Attaourti$^2$ and L. Oufni$^3$\\
{\it {\small $^1$ Universit\'e Sultan Moulay Slimane, FPBM, LIRST,  BP : 523, 23000,  B\'eni Mellal, Morocco. }}\\
 {\it {\small $^2$ Universit\'e Cadi Ayyad, FSSM, LPHEA, BP : 2390, 40000, Marrakech, Morocco }}\\
{\it {\small $^3$ Universit\'e Sultan Moulay Slimane, FSTBM, LPMM,  BP : 523, 23000,  B\'eni Mellal, Morocco. }}}
   \maketitle \setcounter{page}{1}
\begin{abstract}
New insights into our understanding of the  semirelativistic excitation of atomic hydrogen by electronic impact have been made possible by combining the use of polarized electron beams and intense laser field. The paper reviews relativistic theoretical treatment in laser-assisted electron scattering with particular emphasis upon spin effects. Different spin configurations for inelastic electron-atom collisions is also discussed. The role of laser field in such collision is of major importance and reveals new information on the dynamics of the collision process. The examined modern theoretical investigations of such relativistic laser-assisted collisions have shown that the need for experimental data is of a paramount importance in order to asses the accuracy of our calculations.
\vspace{.04cm}\\
PACS number(s): 34.80.Dp, 12.20.Ds
\end{abstract}

\maketitle
\section{Introduction}
The spin is not only an indispensable ingredient in atomic physics but also responsible for many phenomena observed in solid-state physics. In addition to the uses of polarized electrons in studies of atomic physics, there have been numerous studies of polarized electron scattering and polarized
electron emission from ferromagnetic solids over the past decade. In 1975, the purpose of the Spin-Polarized Electron Source was to describe how this effect, which had been discovered in spin-polarized photoemission experiments by Pierce \textit{et al} at the ETH-Zurich \cite{1} , could be used to provide a compact spin-polarized electron gun. Later on, an experimental work has been done to produce electron beams in which the spin has a preferential orientation. They are called polarized electron beams \cite{2} in analogy to polarized light, in which the field vectors have a preferred orientation. Extensive theoretical works have been performed by introducing relativistic and spin effects in the collision between incident particles and atoms [3-6].There are many reasons for the interest in polarized electrons. One important reason is that in physical experiments one endeavors to define as exactly as possible the initial and/or final states of the systems being considered.

Since the 1960s when lasers became a worldwide-used laboratory equipment and also large polarization effects in low-energy
electron scattering were ascertained, experimental and theoretical studies of laser-matter interaction have witnessed continuous progress. By virtue of the increasing progress in the availability of more powerful and tunable lasers , such processes are nowadays being observed in
laboratories [7-10]. Most experimental and theoretical studies of laser-assisted  electron-atom collisions were restricted to the nonrelativistic regime and low-frequency fields, where it has been already recognized that, as a general consequence of the infrared divergence of QED, large numbers of photons can be exchanged between the field and the projectile-target system. An extension of the first-Born nonrelativistic treatment \cite{11} to the relativistic domain was formally derived for unpolarized electrons \cite{12}. There have been also theoretical investigations of relativistic scattering in multimode fields \cite{13}.

In the present paper we have to extended our previous results \cite{14} to the case of laser assisted inelastic excitation $1s$ to $2s$ of the atomic hydrogen by polarized electrons. Therefore, we have begun with the most basic results of our work using atomic units (a.u) in which one has ($\hbar=m_e=e=1$), where $m_e$ is the electron  mass at rest. We have used the metric tensor $g^{\mu\nu}= diag(1,-1,-1,-1)$ and the Lorentz scalar  product defined by $(a.b)=a^\mu b_\mu$. The organization of this paper is as follows : the presentation of the necessary formalism of this work in section  2, the result and discussion in section 3 and at last a brief conclusion in section 4.
\section{Theory}
The transition matrix element corresponding to the laser assisted inelastic excitation of atomic
hydrogen by electronic impact from the initial state $i$ to the final state $f$ is given by
\begin{eqnarray}
S_{fi}=-i\int dt\; \langle \psi_{q_f}(\mathbf{r}_1)\phi_f(\mathbf{r}_2)|V_d| \psi_{q_i}(\mathbf{r}_1)\phi_i(\mathbf{r}_2)\rangle\label{1}
\end{eqnarray}
where $V_d=1/r_{12}-Z/r_1$ is the interaction potential, $\mathbf{r}_1$ are the coordinates of the incident and scattered
electron, $\mathbf{r}_2$ the atomic electron coordinates, $r_{12} = | \mathbf{r}_1 - \mathbf{r}_2 |$ and $r_1 =| \mathbf{r}_1 |$. Before we present the most interesting results of our investigation regarding laser-assisted inelastic excitation of atomic hydrogen by electronic impact, we sketch the principal steps of our theoretical treatment. The solutions of the Dirac equation for an
electron with four-momentum $p^\mu$  inside an electromagnetic plane wave are well known \cite{15}. They read for
the case of circular polarization of the field propagating along the $Oz$ direction
\begin{eqnarray}
\psi_q(x)=\left(1+\frac{\ks\As}{2c(kq)}\right)\frac{u(p,s)}{\sqrt{2VQ_0}}\exp\left[-i(qx)-i\int_0^{kx}\frac{(Ap)}{c(kq)}d\phi\right].\label{2}
\end{eqnarray}
where $u$ represents a free electron bispinor satisfying the Dirac equation without field and which is normalized by  $\overline{u}u=u^\dag\gamma^0u=2c^2$. Here the Feynman slash notation is used, and $V$ is the normalization volume. The physical significance of $q^\mu= (Q/c, \mathbf{q})$ is the averaged four-momentum (dressed momentum) of the particle inside the laser field having a vector potential $A^\mu=(0, a_1 \cos(kx), a_2 \sin(kx), 0)$ with wave four-vector $k^\mu$
: $q^\mu=p^\mu-k^\mu[A^2/2(kp)c^2]$.

In inelastic scattering, it is not only the state of the electron that is changed, but also
the state of the atom. Before starting the calculations, we clarified the different configurations appearing with the orientations of the electron's spin polarizations. We have many  possible scattering scenarios
\begin{center}
\begin{tabular}{|c|c|}
  \hline
 1& $e^{(\uparrow)}   +   H^{(\uparrow)}(1s) \longrightarrow e^{(\uparrow)}   +   H^{(\uparrow)}(2s)$ \\ \hline
2 & $e^{(\downarrow)}  +  H ^{(\uparrow)}(1s) \longrightarrow e^{(\uparrow)} +   H^{(\uparrow)}(2s)$ \\ \hline
 3 & $e^{(\uparrow)} +   H^{(\uparrow)}(1s) \longrightarrow e^{(\downarrow)} +   H^{(\uparrow)}(2s)$ \\ \hline
  4 & $e^{(\uparrow)} +   H^{(\downarrow)}(1s) \longrightarrow e^{(\downarrow)} +   H^{(\downarrow)}(2s)$ \\ \hline
  6 & $e^{(\downarrow)} +   H^{(\downarrow)}(1s) \longrightarrow e^{(\uparrow)} +   H^{(\downarrow)}(2s)$ \\ \hline
 7 & $e^{(\downarrow)} +   H^{(\downarrow)}(1s) \longrightarrow e^{(\downarrow)} +   H^{(\downarrow)}(2s)$\\
  \hline
  \vdots & \vdots\\
  \hline
\end{tabular}
\end{center}
Here, the up and down arrows indicate the direction of the electron's and
atom's spin polarization relative to some fixed axis. During the interaction, the products of states with spin non flip are the same $\varphi_{2s} ^{\dag(\uparrow})(\mathbf{r}_2) \varphi_{1s}^{(\uparrow)}(\mathbf{r}_2)=\varphi_{2s} ^{\dag(\downarrow)}(\mathbf{r}_2) \varphi_{1s}^{(\downarrow)}(\mathbf{r}_2)$ and the product of states with spin flip gives zero
\begin{eqnarray}
\varphi_{2s} ^{\dag(\uparrow)}(\mathbf{r}_2) \varphi_{1s}^{(\downarrow)}(\mathbf{r}_2)&=&\left(
                                                                                 \begin{array}{llll}
                                                                                   2-r_2  ,& 0  ,&  \frac{-i(4-r_2)}{4r_2c}z ,&
                                                                                   \frac{(4-r_2)}{4r_2c}(-y-ix) \\
                                                                                 \end{array}
                                                                               \right)\left(
\begin{array}{c}
 0 \\
  1 \\
  \frac{i}{2cr_2}(x-iy)  \\
-\frac{i}{2cr_2}z
\end{array}\right)\frac{1} {4\sqrt{2}\pi } e^{-2r_{2}}\nonumber\\
&=&\varphi_{2s} ^{\dag(\downarrow)}(\mathbf{r}_2) \varphi_{1s}^{(\uparrow)}(\mathbf{r}_2)\nonumber\\
&=&0\label{3}
\end{eqnarray}
with $\varphi_{2s}(\mathbf{r}_2)$ and $\varphi_{1s}(\mathbf{r}_2)$ are the wave functions of atomic hydrogen corresponding to $2s$ and $1s$ states respectively.
In this case, the probability that the bound electron changes the orientation of its spin in the transition from the initial state $1s$ to the final state $2s$ is zero. Thus, the number of realistic configurations reduces many more.
We review first the basics needed for the description of spin polarization.
Free electrons with four-momentum $p$ and spin $s$ are described
by the free spinors $u(p,s)$, the vector $s^{\mu }$ is defined by
\begin{equation}
s^{\mu }=\frac{1}{c}\left( \left| \mathbf{p}\right| ,\frac{E}{c}\widehat{%
\mathbf{p}}\right),\label{4}
\end{equation}
(with $\widehat{\mathbf{p}}=\mathbf{p}/|\mathbf{p}|$ ) is a
Lorentz vector in a frame in which the particle moves with
momentum $\mathbf{p}$. One easily checks the normalization and the orthogonality conditions respectively
\begin{equation}
s.s=s_{\mu }.s^{\mu }=-1\quad ;\quad p.s=p_{\mu }.s^{\mu }=0.\label{5}
\end{equation}
In practical calculations of quantum electrodynamic (QED) processes, we become acquainted with a technique of calculation which allows the simple treatment of complicated expressions especially the calculation of traces of products of many $\gamma$ matrices. It is based on a projection procedure. The appropriate operators which achieve this are called spin projection operators. In the relativistic case, it is given by
\begin{equation}
\widehat{\Sigma }(s)=\frac{1}{2}(1+\gamma _{5}\ssl),\label{6}
\end{equation}
with $\gamma _{5}=i\gamma ^{0}\gamma ^{1}\gamma ^{2}\gamma
^{3}=-i\gamma _{0}\gamma _{1}\gamma _{2}\gamma _{3}$. This
operator has the following properties
\begin{equation}
\widehat{\Sigma }(s)u(p,\pm s)=\pm  u(p,s).\label{7}
\end{equation}
One can also apply this formalism to helicity states where the
spin points in the direction of the momentum $\mathbf{p}$
\begin{equation}
s_{\lambda }^{\prime}=\lambda \frac{\mathbf{p}}{\left|
\mathbf{p}\right| }=\lambda\widehat{\mathbf{p}}\, \lambda =\pm 1.\label{8}
\end{equation}
We can then define a four spin vector as
\begin{equation}
s_{\lambda }^{\mu }=\frac{\lambda }{c}\left( \left|
\mathbf{p}\right| ,\frac{
E}{c}\widehat{\mathbf{p}}\right).\label{9}
\end{equation}
The starting point of our calculation is the laser-assisted DCS for atomic hydrogen by an electron with well defined momentum $p_{i}$ and
well defined spin $s_{i}$. If the final spin $s_{f}$ is also measured, the polarized DCS then reads as
\begin{eqnarray}
\frac{d\sigma }{d\Omega _{f}}(\lambda_i, \lambda_f)=\sum_{n=-\infty}^{+\infty}\frac{d\sigma^{(n)}}{d\Omega_f}(\lambda_i, \lambda_f),\label{10}
\end{eqnarray}
with
\begin{eqnarray}
\frac{d\sigma^{(n)}}{d\Omega_f}(\lambda_i, \lambda_f)=\left.\frac{|\mathbf{q}_f|}{|\mathbf{q}_i|}\frac{1}{(4\pi
c^2)^2}\left|\overline{u}( p_{f},s_{f})\Gamma_{n}u(p_{i},s_{i})\right|^{2}\left|H_{inel}(\Delta_s)\right|^2\right|_{Q_f=Q_i+n\omega+E_{1s1/2}-E_{2s1/2}}. \label{11}
\end{eqnarray}
The quantity $H_{inel}(\Delta_s)$ which represents the integral part is given by :
\begin{eqnarray}
H_{inel}(\Delta)=-\frac{4\pi}{\sqrt{2}}(I_1+I_2+I_3)
\label{18}
\end{eqnarray}
with $I_1$, $I_2$ and $I_3$ are as follow :
\begin{eqnarray}
I_1&=&\frac{4}{27c^2}\int_0^{+\infty}dr_1\;r_1e^{-\frac{3}{2}r_1}j_0(\Delta r_1)=\frac{4}{27c^2}\frac{1}{((3/2)^2+\mathbf{\Delta}^2)}\nonumber\\
I_2&=&\frac{6}{27}(\frac{1}{c^2}-4)\int_0^{+\infty}dr_1\;r_1^2e^{-\frac{3}{2}r_1}j_0(\Delta r_1)=\frac{2}{27}(\frac{1}{c^2}-4)\frac{3}{((3/2)^2+\mathbf{\Delta}^2)^2}\\
\label{13}
I_3&=&-\frac{4}{9}(1+\frac{1}{8c^2})\int_0^{+\infty}dr_1\;r_1^3e^{-\frac{3}{2}r_1}j_0(\Delta
r_1)=\frac{8}{9}(1+\frac{1}{8c^2})\frac{\mathbf{\Delta}^2-27/4}{((3/2)^2+\mathbf{\Delta}^2)^3}.\nonumber
\end{eqnarray} 
Using REDUCE \cite{17}, the spinorial part obtained after tedious calculations reads as
\begin{eqnarray}
\left|\overline{u}( p_{f},s_{f})\Gamma_{n}u(p_{i},s_{i})\right|^{2}&=&\textbf{Tr}\{ \Gamma _{n}\frac{(1+\lambda _{i}\gamma
_{5}\ssl_{i})}{2}(c\ps_{i}+c^{2})\overline{\Gamma }_{n}\frac{
(1+\lambda _{f}\gamma
_{5}\ssl_{f})}{2}(c\ps_{f}+c^{2})\},\nonumber\\
&=&\big\{J_n^2(z)\mathcal{A}+\big(J^2_{n+1}(z)+J^2_{n-1}(z)\big)\mathcal{B}+\big(J_{n+1}(z)J_{n-1}(z)\big)\mathcal{C}\nonumber\\
&&+J_n(z)\big(J_{n-1}(z)+J_{n+1}(z)\big)\mathcal{D}\big\}.\label{12}
\end{eqnarray}
with $\overline{\Gamma }_{n}=\gamma^0\Gamma _{n}^{\dag}\gamma^0$ and $\Gamma _{n}$ is explicitly detailed in our previous work \cite{18}.\\
Before presenting  our analytical results, we would like to emphasize that the REDUCE code we have written for obtaining the four coefficients $\mathcal{A}$, $\mathcal{B}$, $\mathcal{C}$ and $\mathcal{D}$ gave very long analytical expressions which were difficult to incorporate in the corresponding latex manuscript. Thus, we prefer to give below, for example, just the coefficient $\mathcal{A}$ multiplying the  Bessel function $J_n^2(z)$.
{\small
\begin{eqnarray*}
    \mathcal{A}&=&\frac{1}{(2 (k.p_f)^2 (k.p_i)^2 c^8)}\Big[2 (k.p_f)^2 (k.p_i)^2 \lambda_f \lambda_i |\mathbf{p_f}|^2 |\mathbf{p_i}|^2 c^8 \cos(\theta_{if})-2(k.p_f)^2 (k.p_i)^2 \lambda_f \lambda_i |\mathbf{p_f}|^2 \\
     &&\times  c^6 \cos(\theta_{if}) E_i^2+2 (k.p_f)^2(k.p_i)^2 \lambda_f \lambda_i |\mathbf{p_f}| |\mathbf{p_i}| c^{10}-2 (k.p_f)^2 (k.p_i)^2 \lambda_f \lambda_i |\mathbf{p_i}|^2 c^6 \cos(\theta_{if}) E_f^2\\
     &&+2 (k.p_f)^2 (k.p_i)^2 \lambda_f \lambda_i c^8 \cos(\theta_{if})E_f E_i+2 (k.p_f)^2 (k.p_i)^2 \lambda_f \lambda_i c^4 \cos(\theta_{if}) E_f^2 E_i^2+2 (k.p_f)^2 \\
     &&\times(k.p_i)^2 |\mathbf{p_f}| |\mathbf{p_i}| c^{10} \cos(\theta_{if})+2 (k.p_f)^2 (k.p_i)^2 c^{12}+2 (k.p_f)^2 (k.p_i)^2 c^8 E_f E_i+2 (k.p_f)^2 (k.p_i) \\
      &&\times\lambda_f\lambda_i |a| |\mathbf{p_i}|^2 c^4 \cos(\theta_{if}) E_f \omega-2 (k.p_f)^2 (k.p_i) \lambda_f \lambda_i|a| c^2 \cos(\theta_{if}) E_f E_i^2 \omega-2 (k.p_f)^2 (k.p_i) |a| c^6 \\
      &&\times E_i\omega+2 (k.p_f) (k.p_i)^2 \lambda_f \lambda_i |a| |\mathbf{p_f}|^2 c^4 \cos(\theta_{if}) E_i \omega-2 (k.p_f) (k.p_i)^2 \lambda_f \lambda_i |a| c^2 \cos(\theta_{if}) E_f^2 E_i \omega\\
      &&-2(k.p_f) (k.p_i)^2 |a| c^6 E_f \omega-2 (k.p_f) (k.p_i) (k.s_f) \lambda_f \lambda_i |a||\mathbf{p_f}| |\mathbf{p_i}|^2 c^6 \cos(\theta_{if}) \omega+2 (k.p_f) (k.p_i) \\
     &&\times (k.s_f) \lambda_f \lambda_i |a| |\mathbf{p_f}|c^4 \cos(\theta_{if}) E_i^2 \omega-2 (k.p_f) (k.p_i) (k.s_f) \lambda_f \lambda_i |a| |\mathbf{p_i}| c^8 \omega-2 (k.p_f) (k.p_i) (k.s_i) \\
     &&\times \lambda_f \lambda_i |a| |\mathbf{p_f}|^2 |\mathbf{p_i}| c^6 \cos(\theta_{if})\omega-2 (k.p_f) (k.p_i) (k.s_i) \lambda_f \lambda_i |a| |\mathbf{p_f}| c^8 \omega+2 (k.p_f)(k.p_i) (k.s_i)\lambda_f \lambda_i \\
     &&\times  |a| |\mathbf{p_i}| c^4 \cos(\theta_{if}) E_f^2 \omega-(k.p_f) (k.p_i) \lambda_f\lambda_i a^2 |\mathbf{p_f}| |\mathbf{p_i}| c^2\omega a^2+k.p_f) (k.p_i) \lambda_f \lambda_i a^2 \cos(\theta_{if}) E_f E_i \omega^2\\
      &&+2 (k.p_f)(k.p_i) \lambda_f \lambda_i |a| |\mathbf{p_f}|^2 |\mathbf{p_i}|^2 c^4 \cos(\theta_{if}) \omega^2-2 (k.p_f)(k.p_i) \lambda_f \lambda_i |a| |\mathbf{p_f}|^2 c^2 \cos(\theta_{if}) E_i^2 \omega^2\\
      &&+2 (k.p_f)(k.p_i) \lambda_f \lambda_i |a| |\mathbf{p_f}| |\mathbf{p_i}| c^6 \omega^2-2 (k.p_f) (k.p_i) \lambda_f \lambda_i|a| |\mathbf{p_i}|^2 c^2 \cos(\theta_{if}) E_f^2 \omega^2-2 (k.p_f) (k.p_i)\\
     && \times\lambda_f \lambda_i |a|c^4 \cos(\theta_{if}) E_f E_i \omega^2+2 (k.p_f) (k.p_i) \lambda_f \lambda_i |a| \cos(\theta_{if})E_f^2 E_i^2 \omega^2+(k.p_f) (k.p_i) a^2 c^4 \omega^2\\
     &&-2 (k.p_f)(k.p_i) |a| |\mathbf{p_f}| |\mathbf{p_i}| c^6 \cos(\theta_{if}) \omega^2-2 (k.p_f) (k.p_i) |a| c^8 \omega^2+2 (k.p_f) (k.p_i) |a| c^4 E_f E_i \omega^2\\
     &&+(k.p_f) (k.s_i) \lambda_f\lambda_i a^2 |\mathbf{p_f}| c^2 E_i \omega^2-(k.p_f) (k.s_i) \lambda_f \lambda_i a^2 |\mathbf{p_i}|c^2 \cos(\theta_{if}) E_f \omega^2-(k.p_i)(k.s_f) \lambda_f \lambda_i\\
     && \times  a^2 |\mathbf{p_f}| c^2 \cos(\theta_{if})E_i \omega^2+(k.p_i) (k.s_f) \lambda_f \lambda_i a^2 |\mathbf{p_i}| c^2 E_f \omega^2+(k.s_f) (k.s_i) \lambda_f \lambda_i a^2 |\mathbf{p_f}| |\mathbf{p_i}| c^4 \cos(\theta_{if}) \\
    &&\times\omega^2+(k.s_f) (k.s_i)\lambda_f \lambda_i a^2 c^6 \omega^2-(k.s_f) (k.s_i) \lambda_f \lambda_i a^2 c^2 E_f E_i \omega^2\Big]
\end{eqnarray*}
}
At this stage, note that if $\lambda_i \lambda_f=1$ during the scattering process, which physically means that there is no helicity flip occurring but if $\lambda_i \lambda_f=-1$, this means that a helicity flip occurred. In the absence of the laser field ($|a|=0)$, this coefficient reduces to
{\small
\begin{eqnarray*}
\mathcal{A}&=&\frac{1}{c^4}\Big[\lambda_f \lambda_i |\mathbf{p_f}|^2 |\mathbf{p_i}|^2 c^4 \cos(\theta_{if})-\lambda_f \lambda_i |\mathbf{p_f}|^2 c^2 \cos(\theta_{if})
     E_i^2+\lambda_f \lambda_i |\mathbf{p_f}| |\mathbf{p_i}| c^6-\lambda_f \lambda_i |\mathbf{p_i}|^2 c^2 \cos(\theta_{if}) E_f^2\\
    && +\lambda_f \lambda_i c^4 \cos(\theta_{if}) E_f E_i+\lambda_f \lambda_i \cos(\theta_{if}) E_f^2 E_i^2+|\mathbf{p_f}|
     |\mathbf{p_i}| c^6 \cos(\theta_{if})+c^8+c^4 E_f E_i\Big]
\end{eqnarray*}
}
We note that without laser field, equations (\ref{10})
and (\ref{11}) reduce to the well-known polarized (first-Born) differential cross section of inelastic excitation of atomic hydrogen by electronic impact.
\begin{figure}[!h]
\centering
\includegraphics[angle=0,width=3in,height=3.5 in]{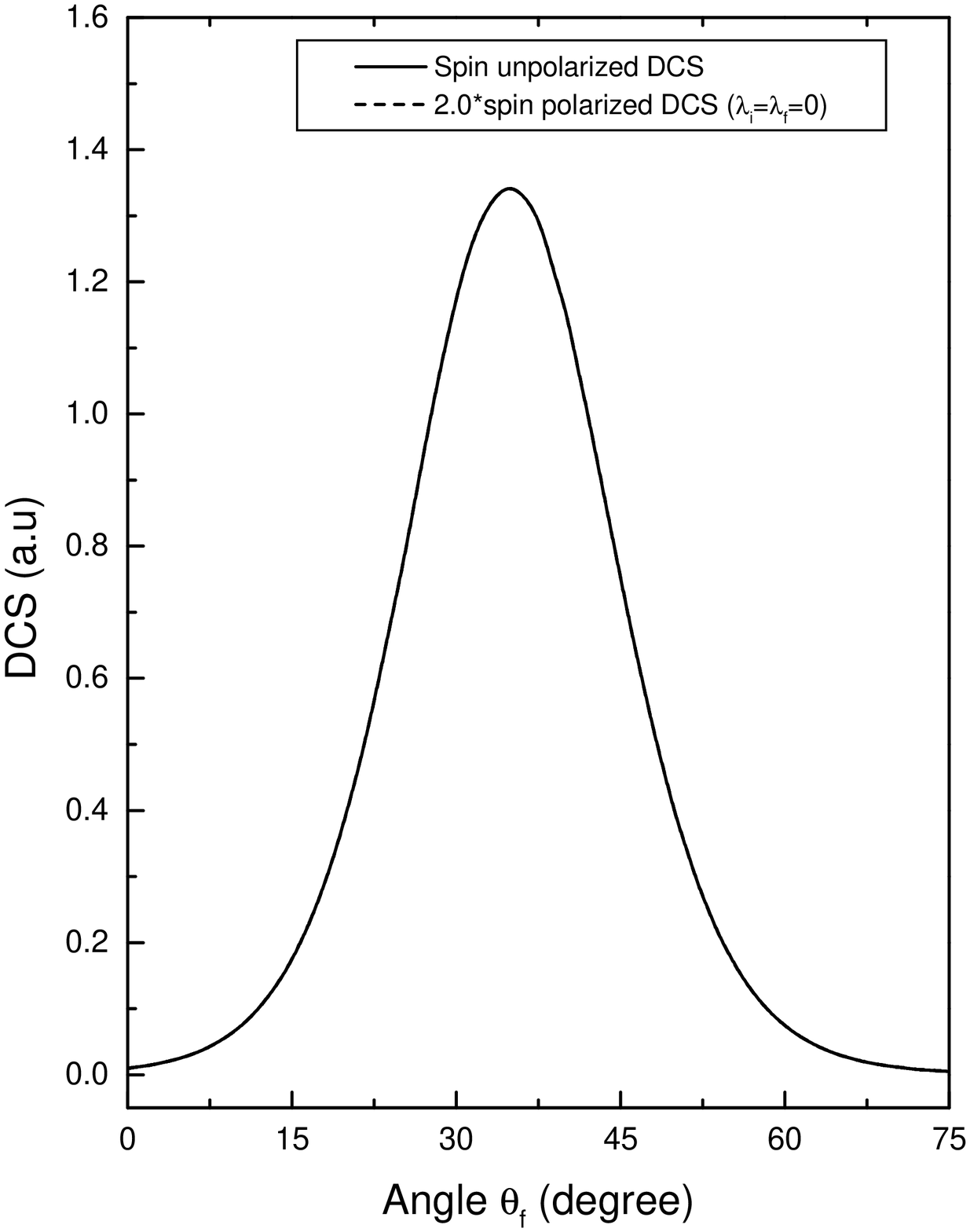}
\caption{The different TDCSs (Unpolarized DCS, Spin polarized DCS with ($\lambda_i=\lambda_f=0$)) scaled in $10^{-9}$ as a function of the angle $\theta_f$ . The relativistic parameter is $\gamma=1.0053$, the electrical field strength is $\mathcal{E}=0.05\;a.u$. The geometric parameters are $\theta_i=45^{\circ}$, $\phi_i=0^{\circ}$,  $\phi_f=45^{\circ}$ and the number of photons exchanged are $n=\pm 100$.}
\end{figure}
\section{Results and discussions}
The spin-polarized differential cross  sections $ d\sigma^{(n)}/d\Omega_f$ are computed for the laser-assisted semirelativistic excitation of atomic hydrogen by electronic impact, where $n$ denotes the number of photons absorbed or emitted. With a
view to qualitative comparison with our previous work \cite{14}, kinematics and the geometry parameters are chosen in accordance with those used in \cite{14}. The direction of the laser field is chosen parallel to $Oz$ axis. The corresponding spin unpolarized differential cross section $ d\overline{\sigma}^{(n)}/d\Omega_f$ results are also presented for comparison.
\begin{figure}[!h]
\centering
\includegraphics[angle=0,width=3in,height=3.5 in]{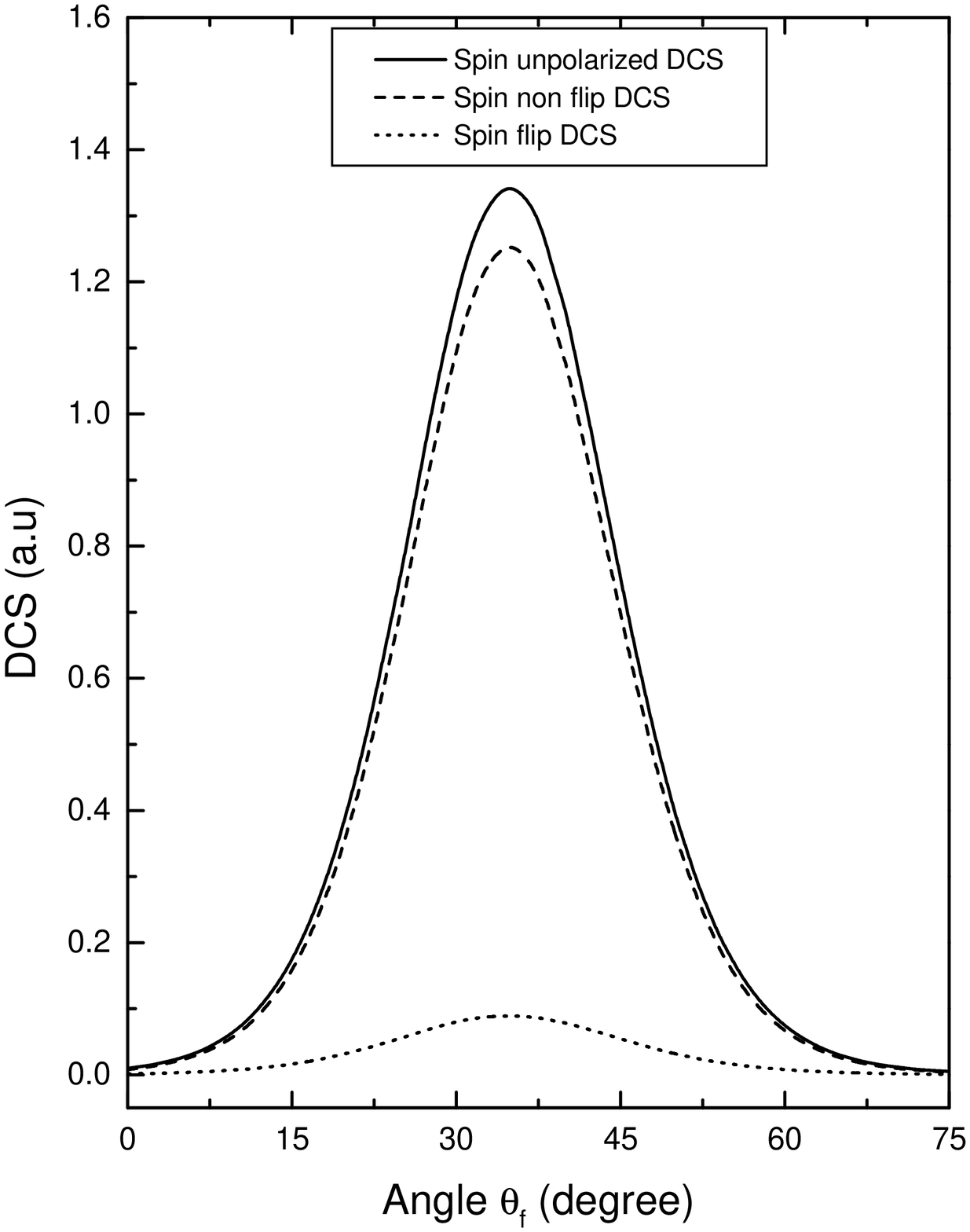}
\caption{The different TDCSs (Unpolarized DCS, Spin flip DCS and Spin non flip DCS) scaled in $10^{-9}$ as a function of the angle $\theta_f$ . The relativistic parameter is $\gamma=1.0053$, the electrical field strength is $\mathcal{E}=0.05\;a.u$. The geometric parameters are $\theta_i=45^{\circ}$, $\phi_i=0^{\circ}$,  $\phi_f=45^{\circ}$ and the number of photons exchanged are $n=\pm 100$.}
\end{figure}
In view of equation (\ref{12}) if the mathematical condition ($\lambda_i=\lambda_f=0$) is used, the spinorial part takes the following form
\begin{eqnarray}
\left|\overline{u}( p_{f},s_{f})\Gamma_{n}u(p_{i},s_{i})\right|^{2}=\frac{1}{4}\textbf{Tr}\{ \Gamma _{n}(c\ps_{i}+c^{2})\overline{\Gamma }_{n}(c\ps_{f}+c^{2})\}\label{13}
\end{eqnarray}
In comparison with the unpolarized spinorial part
\begin{eqnarray}
\frac{1}{2}\sum_{s_is_f}\left|\overline{u}( p_{f},s_{f})\Gamma_{n}u(p_{i},s_{i})\right|^{2}=\textbf{Tr}\{ \Gamma _{n}(c\ps_{i}+c^{2})\overline{\Gamma }_{n}(c\ps_{f}+c^{2})\}\label{14}
\end{eqnarray}
it may be noted from equations (\ref{13}) and  (\ref{14}) that the spin unpolarized DCS is equal to two times to the spin polarized DCS but only under the condition ($\lambda_i=\lambda_f=0$)
\begin{eqnarray}
\frac{d\overline{\sigma}^{(n)} }{d\Omega _{f}}=2\times \frac{d\sigma^{(n)}}{d\Omega_f}(\lambda_i=0, \lambda_f=0),\label{15}
\end{eqnarray}
this equation represents our first consistency check. The numerical results  are displayed in figure 1, where we have plotted the two DCSs (2 times spin polarized DCS with ($\lambda_i=\lambda_f=0$) and spin unpolarized DCS ) by varying the angle $\theta_f$ of the scattered electron. As it is seen in this figure, we have indistinguishable curves. The most important result (second consistency check) is the sum of  polarized DCS (spin flip) and  polarized DCS (spin non flip) always gives the spin unpolarized DCS as this is shown in figure 2.\\
\begin{figure}[!h]
\centering
\includegraphics[angle=0,width=4in,height=4 in]{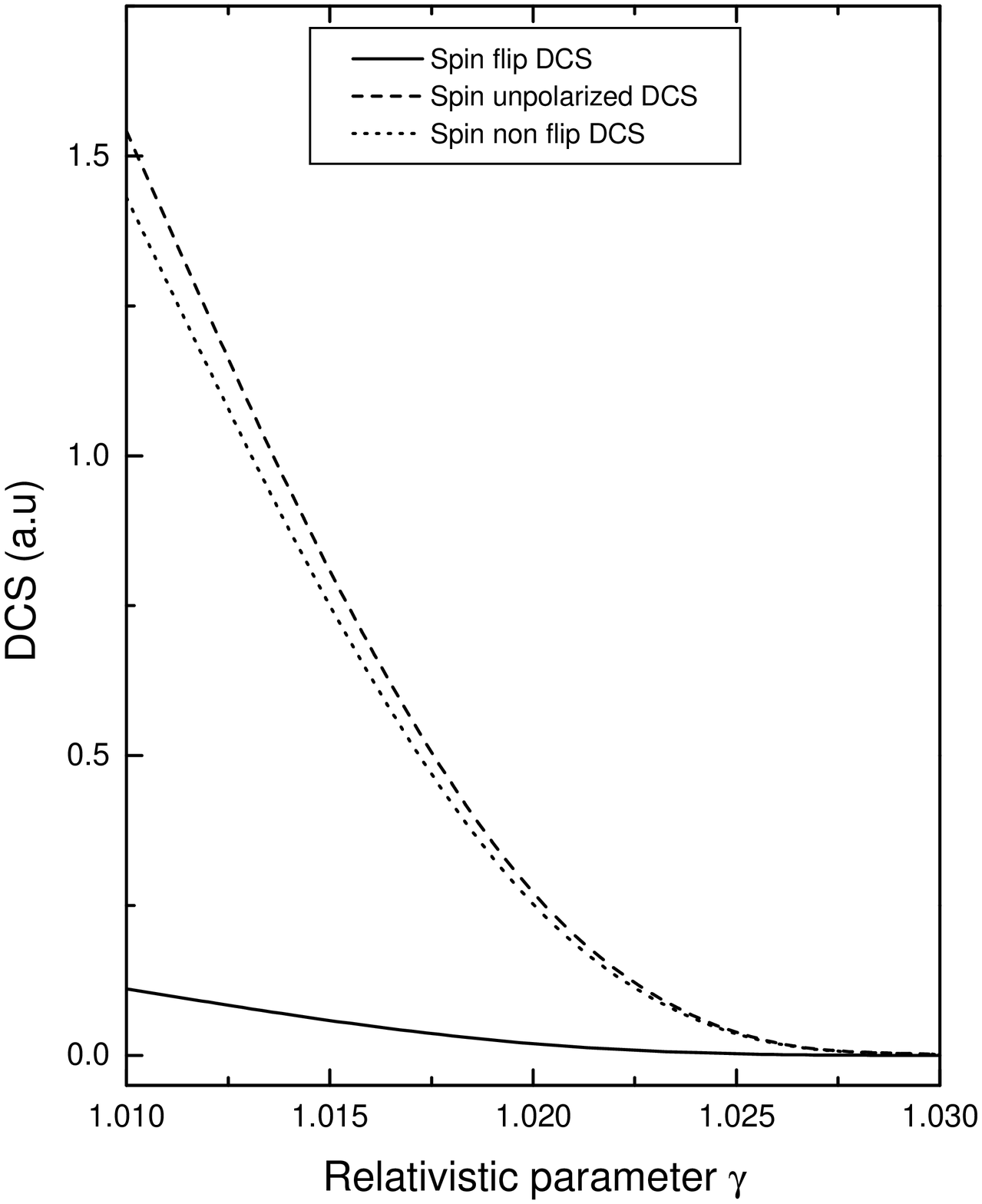}
\caption{The different TDCSs (Unpolarized DCS, Spin flip DCS and Spin non flip DCS) scaled in $10^{-13}$ as a function of the relativistic parameter $\gamma$, the geometric parameters are $\theta_i=45^{\circ}$, $\phi_i=0^{\circ}$,  $\phi_f=45^{\circ}$ and $\theta_f=45^{\circ}$.  The electrical field strength is $\mathcal{E}=0.5\;a.u$ and the number of photons exchanged are $n=\pm 50$. }
\end{figure}
Figure 3 shows the different spin polarized and unpolarized DCSs versus the relativistic parameter $\gamma$. It is apparent from
this figure that the kinetic energy of the incident electron has an important effect in the spin orientation. For much of the high energy range,
 the spin polarized  DCS (spin non flip) is approximately equal to the spin unpolarized DCS and the spin polarized DCS (spin flip) is equal to zero.
 The physical meaning of this result is that : in high energy, the probability that the incident electron changes its spin is zero.\\
\begin{figure}[!h]
\centering
\includegraphics[angle=0,width=4in,height=4 in]{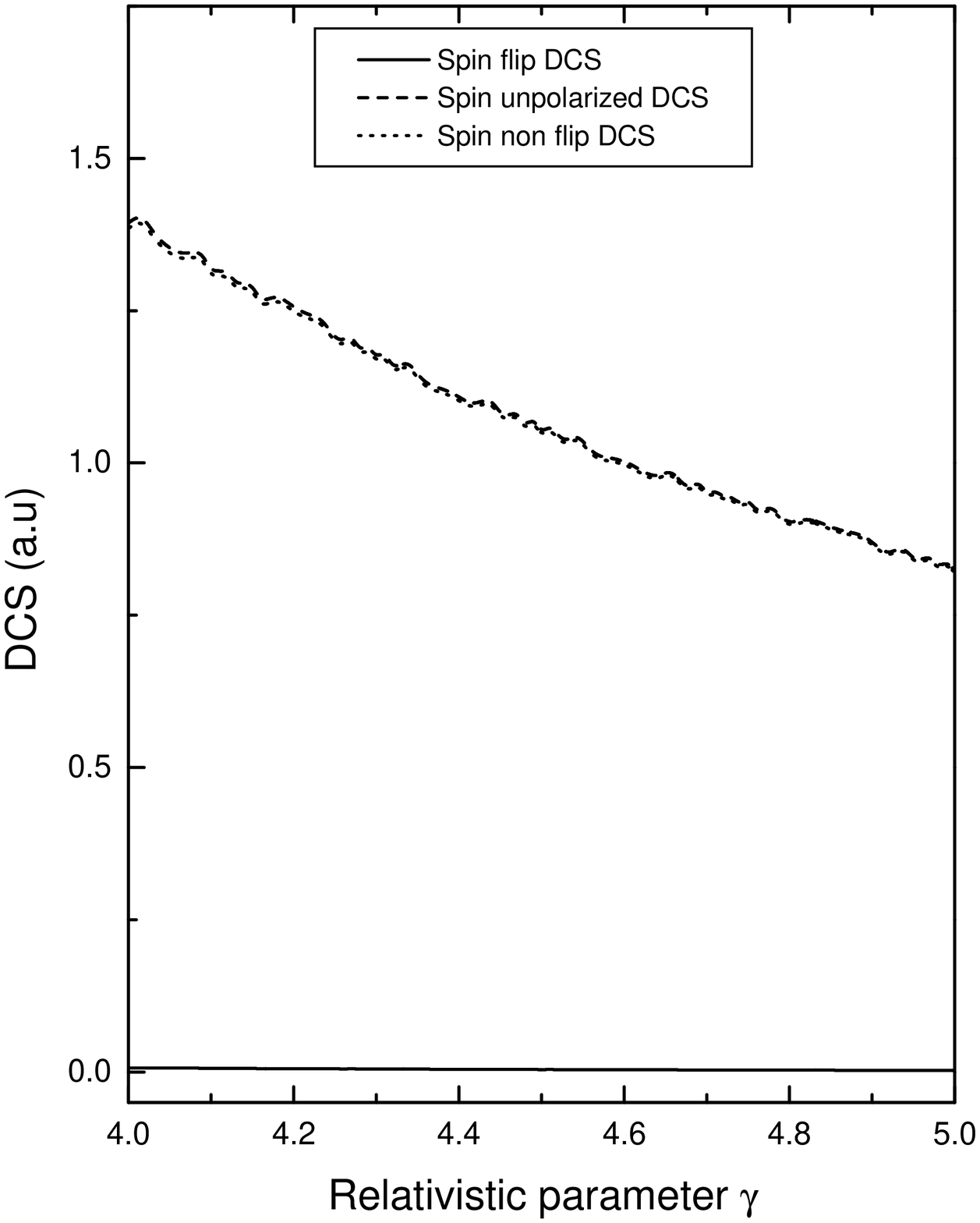}
\caption{The different TDCSs (Unpolarized DCS, Spin flip DCS and Spin non flip DCS) scaled in $10^{-22}$ as a function of the relativistic parameter $\gamma$, the geometric parameters are $\theta_i=45^{\circ}$, $\phi_i=0^{\circ}$,  $\phi_f=45^{\circ}$ and $\theta_f=45^{\circ}$. The electrical field strength is $\mathcal{E}=0.5\;a.u$ and the number of photons exchanged are $n=\pm 50$.}
\end{figure}
In order to clarify the situation in which we have seemingly overlapping curves for the three approaches in figure 3, we give in figure 4 the three approaches (spin polarized (spin flip and spin non flip) and the spin polarized DCS). As it is noticed, the spin polarized DCS (spin non flip) and the spin unpolarized DCS overlap but the spin polarized DCS (spin flip) converges to zero.\\
\begin{figure}[!h]
\centering
\includegraphics[angle=0,width=4in,height=4 in]{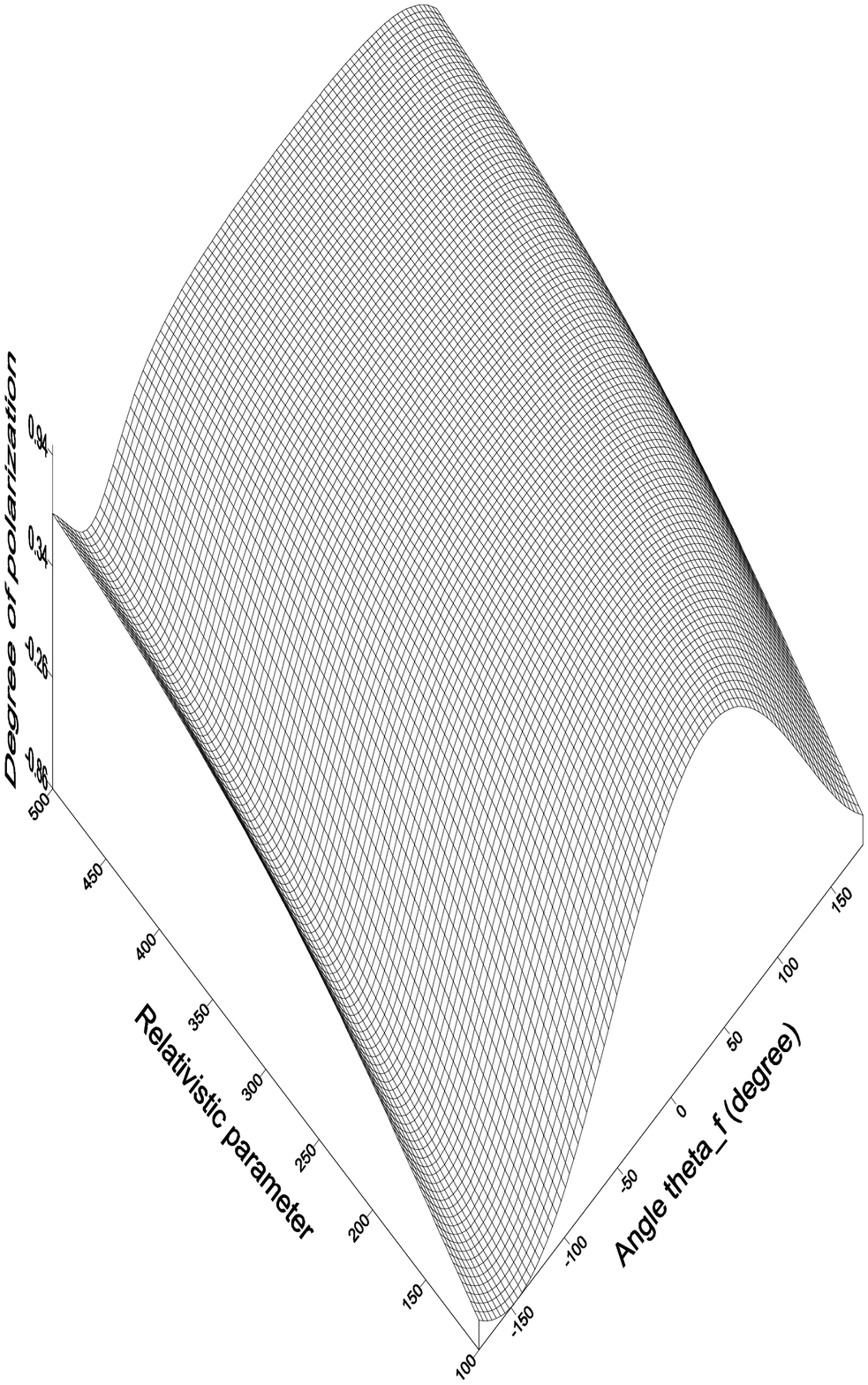}
\caption{The behavior of the degree of polarization $P$ as a function
of the angle $\theta_f$ varying from $-180^{\circ}$ to $180^{\circ}$ and the relativistic
parameter $\gamma$ scaled in $10^{-2}$ in absence of the laser field.}
\end{figure}
Our main interest is the polarization degree of the electron after the event. This quantity is defined as
\begin{eqnarray}
P=\frac{\frac{d\sigma^{(n)} }{d\Omega _{f}}(\textit{spin non flip})-\frac{d\sigma^{(n)}}{d\Omega_f}(\textit{spin flip})}{\frac{d\sigma^{(n)} }{d\Omega _{f}}(\textit{spin non flip})+\frac{d\sigma^{(n)}}{d\Omega_f}(\textit{spin flip})},\label{16}
\end{eqnarray}
figure 5 shows the polarization degree, which is related by
equation (\ref{16}) to the spin polarized differential cross section ratio. A tree dimensional plot of this quantity is given versus the angle $\theta_f$ and the relativistic parameter $\gamma$. The first observation that can be made concerns
the shape of the polarization degree that is strongly changed with
the relativistic parameter $\gamma$. An interesting behavior emerges with increasing
$\gamma$, particularly for $\theta_f=0^{\circ}$, where the polarization degree
performs a plateau-like behavior. This
emphasizes the fact that the spin polarized DCS is very
sensitive to the variation of the relativistic parameter $\gamma$ and this fact has to remain true
for the case in the presence of the laser field. This is consistent with the results shown in figure 3 and 4.
\section{Conclusion}
We have studied the laser assisted inelastic excitation of atomic hydrogen by polarized electrons. We unraveled the influence of the orientation of the spin-polarization of the incoming and scattered electrons relative to the orientation of the spin-polarization of the bound electron. These spin effects depend strongly on the energies of the incoming electron. It is observed that in the transition from $1s$ state to $2s$ state, the electron's probability for changing its spin is zero.

\end{document}